# DNA Sequencing via Quantum Mechanics and Machine Learning

Henry Yuen[1], Fuyuki Shimojo[1,2], Kevin J. Zhang[1,3], Ken-ichi Nomura[1],
Rajiv K. Kalia[1], Aiichiro Nakano[1,1], and Priya Vashishta[1]

[1] Collaboratory for Advanced Computing and Simulations,
Department of Computer Science, Department of Physics & Astronomy,
Department of Chemical Engineering & Materials Science,
University of Southern California, Los Angeles, CA 90089-0242, USA
(hyuen, knomura, rkalia, anakano, priyav)@usc.edu

[2] Department of Physics, Kumamoto University, Kumamoto 860-8555, Japan
shimojo@kumamoto-u.ac.jp

[3] Harvard University, Cambridge, MA 02138
kzhang@college.harvard.edu

**Abstract**. Rapid sequencing of individual human genome is prerequisite to genomic medicine, where diseases will be prevented by preemptive cures. Quantum-mechanical tunneling through single-stranded DNA in a solid-state nanopore has been proposed for rapid DNA sequencing, but unfortunately the tunneling current alone cannot distinguish the four nucleotides due to large fluctuations in molecular conformation and solvent. Here, we propose a machine-learning approach applied to the tunneling current-voltage (I-V) characteristic for efficient discrimination between the four nucleotides. We first combine principal component analysis (PCA) and fuzzy c-means (FCM) clustering to learn the "fingerprints" of the electronic density-of-states (DOS) of the four nucleotides, which can be derived from the I-V data. We then apply the hidden Markov model and the Viterbi algorithm to sequence a time series of DOS data (*i.e.*, to solve the sequencing problem). Numerical experiments show that the PCA-FCM approach can classify unlabeled DOS data with 91% accuracy. Furthermore, the classification is found to be robust against moderate levels of noise, *i.e.*, 70% accuracy is retained with a signal-to-noise ratio of −26 dB. The PCA-FCM-Viterbi approach provides a 4-fold increase in accuracy for the sequencing problem compared with PCA alone. In conjunction with recent developments in nanotechnology, this machine-learning method may pave the way to the much-awaited rapid, low-cost genome sequencer.

---

[1] Corresponding Author. Fax: +1 (213) 821-2664, Email: anakano@usc.edu.

# 1 Introduction

DNA sequencing determines the order of four nucleotide bases—adenine (A), guanine (G), cytosine (C), and thymine (T)—that constitute a DNA molecule. State-of-the-art methods for DNA sequencing, while much improved from the technology used to first sequence the human genome [1, 2], are still costly and time consuming. Today individuals can get their full genome sequenced for about 50,000 USD [3]. Lower-cost, rapid sequencing of individual DNA would enable genomic medicine, where diseases are prevented by preemptive cures. Consequently, the quest for the so-called "ultra-low cost sequencer" (ULCS) is motivating much research into alternative ways of genome sequencing.

One of the active avenues toward ULCS is nanopore sequencing, which infers the base sequence by probing the changes in certain physical signals as the DNA strand threads through a ~2 nm nanopore [4]. Deamer and Akeson survey prospects for different approaches to nanopore sequencing in Ref. 5. While traditional sequencing methods usually require massive replication of the DNA, a nanopore sequencer in principle requires only a single strand. Possibility of nanopore DNA sequencing has been studied by electrically driving a DNA molecule in an electrolyte solution through a biological nanopore and then measuring the changes in ionic current through the nanopore induced by blockaded ions [6, 7]. For example, Ashkenasy *et al*. have investigated single-nucleotide identification capabilities of α-hemolysin nanopore sequencers in Ref. 8. Other groups have tried alternative approaches such as using exonuclease enzymes to cleave individual nucleotides, which are fed to a detection system in order [9]. However, the poor signal-to-noise ratio and the stringent environmental conditions of these schemes have motivated other groups to explore solid-state alternatives involving silicon-based materials [10, 11]. Advantages of using solid-state nanopores include a wider range of operable environmental conditions and the ability to embed sensors directly onto the pore [12].

In spite of the potential of solid-state nanopores for DNA sequencing, distinguishing between the four bases (A, G, C, and T) based on the ionic current through the nanopores remains difficult. This has led several groups to instead measure the transverse electronic tunneling current between electrodes attached to the nanopore. Jauregui *et al*. have performed first-principles calculations to indicate the feasibility of transverse current-based sequencing [13], whereas Gracheva *et al*. have suggested the use of MOS capacitor membranes for constructing similar DNA sequencers [14]. Lagerqvist *et al*. have extended this idea further by proposing to analyze the distribution of transverse current values for each nucleotide as the DNA molecule translocates through the nanopore [15]. This can be accomplished by slowing the speed of DNA translocation, allowing the device to take multiple measurements of the same nucleotide. Though the distributions reveal more about the structure of the passing nucleotide, the distributions of the four nucleotides still show high degrees of overlap and thus the sequencing process is prone to large errors.

The main contribution of this paper is an algorithm, which applies machine-learning techniques to quantum-mechanical tunneling current, in order to sequence single-stranded DNA. Our approach mitigates the problem of indistinguishability of the DNA bases as well as that of noise. Here, we distinguish the nucleotides using their electronic-structure information in the form of the electronic density of states (DOS), which can be acquired by a nanopore device by capturing electronic tunneling current values flowing across the nanopore diameter over a range of voltages. The DOS can then be computed from the derivative of the resulting current-voltage (I-V) curve. In fact, several groups have measured the DOS of DNA using scanning tunneling microscopy [16-18], and one group has conducted theoretical investigations of nanopore-based DOS measurements of DNA molecules [19]. Though the DOS provides the "fingerprints" of the four nucleotides, the I-V characteristics still suffer from large noises due to fluctuations in molecular conformation and solvent (*i.e.*, water molecules and ions). In this paper, we introduce a machine-learning approach to address this problem. We first use principal component analysis (PCA) [20] to algorithmically learn the distinguishing features between the DOS between the four bases, and use these features to predict the identities of an unknown DOS. PCA allows the projection of DOS onto a small-dimensional feature space, in which the four DNA bases form highly disjoint clusters. Then, a given DOS is classified relatively easily [21] by employing the fuzzy c-means (FCM) clustering approach [22]. Finally, we take advantage of the similarity of the DNA sequencing problem to speech recognition and employ the hidden Markov model [23] and the Viterbi algorithm [24] to determine the most likely base sequence from noisy observation. Our numerical experiments demonstrate that the accuracy of nanopore sequencing is significantly improved by the PCA-FCM-Viterbi approach.

This paper is organized as follows: Section 2 presents the PCA-FCM-Viterbi algorithm. Performance of the algorithm is evaluated in section 3, and conclusions are drawn in section 4.

## 2 Method

In this section, we first formulate the DNA sequencing problem in the context of tunneling current-based nanopore sequencing. We then present our PCA-FCM-Viterbi approach to the sequencing problem.

### 2.1 Statement of the Problem

Let $(s_1,...,s_n) \in \{A,T,G,C\}^n$ be a sequence of $n$ bases in single-stranded DNA (ssDNA), where $s_1$ is the base that is about to enter the nanopore sequencer at time $t = 0$ (see Fig. 1). The ssDNA strand will translocate through the nanopore between the electrodes, and at each time $t$ (we measure the time in unit of a time discretization unit, $\Delta t$), the device measures the transverse current (labeled $I$ in the figure) over a range of voltage $V$, thereby generating an $I$-$V$ plot at each time $t$, denoted as the function $I_t(V)$. We then compute $dI_t/dV$, which is proportional to the electronic density of states (DOS) of the group of atoms near the electrodes at time $t$ [16-18]. Let $D_t(E)$ be the DOS as a function of the electron energy $E$ at time $t$. For all time steps $0 \leq t \leq T$ during the measurement, we discretize $D_t(E)$ by creating a vector $\mathbf{D}(t) = (v_1, \ldots, v_B)$, with $v_i = \int_{a+(i-1)\sigma}^{a+i\sigma} D_t(E)dE$. Namely, the continuous function $D_t(E)$ is transformed into a discrete histogram in the energy range $[E_{\min}, E_{\max}]$ with $B$ bins of width $\sigma = (E_{\min} - E_{\max})/B$ (see Fig. 1). Our measurement data is thus a time-series of $B$-dimensional vectors, $(\mathbf{D}(t) \mid 0 \leq t \leq T)$. We would like to point out that generally, $n \ll T$, to provide a sufficiently large number of measurements per nucleotides.

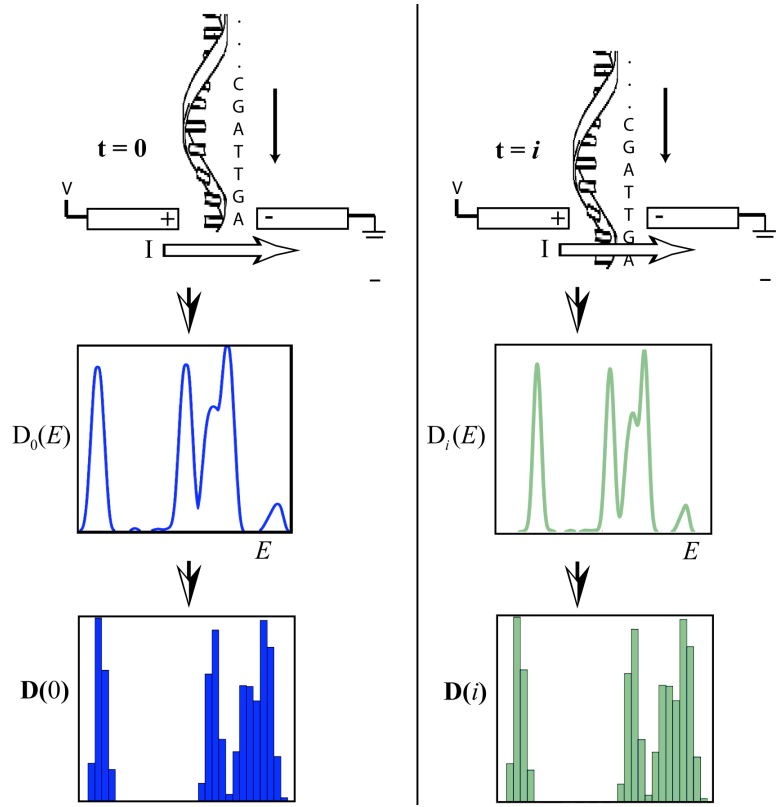

**Fig. 1.** Schematic of traversal tunneling current-based sequencing of a single-stranded DNA translocating through a solid-state nanopore, where $V$ is the applied voltage, and the arrow next to the ssDNA indicates the direction of translocation. At each time step $t$, the device measures the induced transverse current $I$ (the horizontal open arrow) across the electrodes (labeled $+$ and $-$) as a function of $V$, which is varied. We then compute $dI_t/dV$, which is proportional to the electronic density of states of the group of atoms near the electrodes at time $t$ (denoted $D_i(E)$ in the figure). Finally, the continuous DOS function $D_i(E)$ is discretized into a $B$-dimensional vector $\mathbf{D}(t)$, and $(\mathbf{D}(t) \mid 0 \leq t \leq T)$ forms our time series. The figure shows the positions of an ssDNA molecule at two different times, $t = 0$ and $i$, and the double-shaded arrows demonstrate the transformation of data—from physical measurement of the I-V relation to the continuous DOS function $D_i(E)$ to the discretized DOS histogram $\mathbf{D}(t)$, the last of which forms an input to our machine-learning algorithm. The algorithm will analyze the time series $(\mathbf{D}(t) \mid 0 \leq t \leq T)$ to infer the original sequence of bases.

The problem is then stated as follows: Given a time sequence of DOS histograms, $(\mathbf{D}(t) \mid 0 \leq t \leq T)$, determine the original sequence of bases $(s_1,...,s_n)$. In this paper, we break this problem into three sub-problems:

1. *Learning the identifying features of the nucleotide density of states.* First, we algorithmically extract the salient features of the DOS histograms from a training dataset. For this task, we use principal component analysis (PCA), which is a method of statistical data analysis that non-parametrically performs dimension reduction on complex datasets [22]. Applying PCA to the training set will yield a set of principal component (PC) vectors that serve as a basis for representing the original dataset. The PCs, however, are ordered in such a fashion that the first few basis vectors "explain" the variance in the data. That is, the projection of the original dataset on the first few basis vectors has the maximum amount of variance.

2. *Classifying an unlabeled density of states.* With the PC vectors, we now attempt to compute the correct label for a given DOS histogram. We use the membership formula from the fuzzy c-means algorithm to give probabilities that the DOS histogram is a measurement of a nucleotide of type A, T, C, or G.

3. *Sequencing a time series of density of states.* A nanopore sequencing device provides a time series, $(\mathbf{D}(t) \mid 0 \leq t \leq T)$, of DOS, but each $\mathbf{D}(t)$ does not necessarily specify a single nucleotide. In fact, for many $t$, $\mathbf{D}(t)$ is a mixture of DOS's of consecutive nucleotides. Therefore, simply running the classification algorithm on each $\mathbf{D}(t)$ will not work—the identity of $\mathbf{D}(t)$ might depend on the identities of $\mathbf{D}(u < t)$ and $\mathbf{D}(v > t)$. We propose to model the sequence $(\mathbf{D}(t) \mid 0 \leq t \leq T)$ as the output of a hidden Markov model, and use the Viterbi algorithm to determine the most likely sequence of nucleotides underlying $\mathbf{D}(t)$.

## 2.2 Learning the Salient Features of Density of States

We use principal component analysis (PCA) to extract principal components (PCs) from a training set of $4m$ histograms ($m$ histograms per DNA base). We refer the reader to [21], which provides an excellent tutorial for the PCA algorithm. The $4m$ histograms are organized into a $4m \times B$ matrix $\mathbf{H}$, where the matrix element $\mathrm{H}_{ij}$ corresponds to the $j^{\text{th}}$ bin of the $i^{\text{th}}$ histogram ($B$ is the dimension of the DOS vector defined in section 2.1). PCA is then performed in the following manner: From $\mathbf{H}$ construct a new matrix $\mathbf{X} = (\mathbf{H} - \mathbf{M}) / (4m)^{1/2}$, where the column $j$ of $\mathbf{M}$ is $(\mu_j,...,\mu_j)$ with $\mu_j$ the mean of column $j$ of $\mathbf{H}$. Thus, each column of $\mathbf{X}$ has zero mean. Perform singular value decomposition on the matrix $\mathbf{X} = \mathbf{U}\boldsymbol{\Sigma}\mathbf{V}^T$, with the singular values of $\mathbf{X}$ in descending order along the diagonal of $\boldsymbol{\Sigma}$, where $\mathbf{V}$ is a $N \times N$ unitary matrix such that the columns of $\mathbf{V}$ form an orthonormal basis for $\mathbf{X}$. Let $\mathbf{p}_1, \mathbf{p}_2, ..., \mathbf{p}_B$ be the columns of $\mathbf{V}$. According to the theory of PCA, $\mathbf{p}_1$ is the axis along which the variance of $\mathbf{X}$ is maximized, $\mathbf{p}_2$ is an axis orthogonal to $\mathbf{p}_1$ along which the variance of $\mathbf{X}$ is maximized in the subspace orthogonal to $\mathbf{p}_1$, and so on. We call $\mathbf{p}_1, \mathbf{p}_2, ..., \mathbf{p}_B$ the PCs of $\mathbf{X}$.

Let $\mathbf{d}$ be one of the $4m$ histograms. We project $\mathbf{d}$ onto $\mathbf{p}_1$ and $\mathbf{p}_2$, and then over all $\mathbf{d}$, we collect the coordinate pairs $(\mathbf{d}^T\mathbf{p}_1, \mathbf{d}^T\mathbf{p}_2)$ to form the projection space of the dataset, where $\mathbf{d}^T$ denotes the transpose of vector $\mathbf{d}$. The projections of the $m$ histograms of type A, T, C, or G form a cluster of points with centers $\mathbf{c}_A$, $\mathbf{c}_T$, $\mathbf{c}_C$ and $\mathbf{c}_G$, respectively.

## 2.3  Classifying Density of States

To classify an unknown DOS vector $\mathbf{d}$, we project it on $\mathbf{p}_1$ and $\mathbf{p}_2$ and compare the coordinate pair $(\mathbf{d}^T\mathbf{p}_1, \mathbf{d}^T\mathbf{p}_2)$ with the clusters found with the training set. To do so, we construct a set of membership functions, derived from the fuzzy c-means (FCM) clustering algorithm [22]:

$$p_X(\mathbf{d}) = \left( \sum_{N \in \{A,T,C,G\}} \frac{\left\| \mathbf{d}^T\mathbf{p} - \mathbf{c}_x \right\|}{\left\| \mathbf{d}^T\mathbf{p} - \mathbf{c}_N \right\|} \right)^{-1}, \tag{1}$$

where $X$ is one of $\{A,T,C,G\}$, $\mathbf{c}_x$ is the center of the cluster corresponding to nucleotide $X$, and $\mathbf{p}$ is the $m \times 2$ matrix whose columns are $\mathbf{p}_1$ and $\mathbf{p}_2$. $p_X$ takes a $1 \times m$ input histogram $\mathbf{d}$, and returns a real number in $[0,1]$ that represents the degree of membership that histogram $\mathbf{d}$ has in cluster $X$. We can interpret this as the probability that histogram $\mathbf{d}$ comes from a nucleotide of type $X$.

## 2.4  Sequencing a Time Series of Density of States

As mentioned earlier, sequencing a time series of DOS's requires not only identifying a DOS vector $\mathbf{D}(t)$ alone, but possibly $\mathbf{D}(t)$ in conjunction with nearby $\mathbf{D}(u < t)$ or $\mathbf{D}(v > t)$. We make the following simplifying assumptions:

1. The detection range of the electrodes is not much larger than 2-3 nucleotides (6 to 9 Å); then it is possible to have the device measure the local DOS of a single nucleotide. That is, if at time $t$ an A nucleotide on the DNA strand were directly centered between the electrodes, then $\mathbf{D}(t)$ would be nearly identical to the DOS of an isolated A nucleotide.

2. The translocation speed of the DNA strand is uniform.

3. The translocation speed of the DNA strand is high enough such that after 2-3 time steps, the contribution of the nucleotide measured at time $t$ to the DOS measurement at time $t + 2$ (or $t + 3$) is negligible.

These assumptions enforce the "locality" of measurement: The input to our algorithm is a time series $(\mathbf{D}(t) \mid 0 \leq t \leq T)$ such that the identity of $\mathbf{D}(t_i)$ depends at most on $(\mathbf{D}(t_i + k) \mid -2 \leq k \leq 2)$. In addition, our assumptions imply that if $\mathbf{D}(t_i)$ is a pure DOS of a nucleotide that is directly being probed, then at most two consecutive DOS's (*i.e.* $\mathbf{D}(t_i + 1)$ and $\mathbf{D}(t_i + 2)$) will be mixtures of DOS's of consecutive bases, before the next DOS ($\mathbf{D}(t_i + 3)$) will again be a pure measurement of a nucleotide.

We compare two approaches to the time-series sequencing problem. First, we attempt a naive approach to sequence $\mathbf{D}(t)$ by simply computing $p_X(\mathbf{D}(t))$ for each $X$ in $\{A, T, C, G\}$, at each $t$, and identify the measurement $\mathbf{D}(t)$ with the $X$ such that $p_X(\mathbf{D}(t)) \geq \rho$ where $\rho$ is a probability threshold greater than, *e.g.*, 0.5. If there is no such $X$, then we take this to mean that $\mathbf{D}(t)$ corresponds to a mixed DOS of more than one nucleotides, and we simply discard this measurement. We take the sequence of decoded symbols that are left as the estimated sequence of bases. The following subsections describe the second approach based on the hidden Markov model and the Viterbi algorithm.

### 2.4.1  Hidden Markov Model

In the second approach, we model the process of measurement generation with the hidden Markov model (HMM) formalism, in which we use the Viterbi algorithm to calculate the most likely sequence of nucleotides that generated the given sequence of observations. HMMs models sequential, stochastic processes where at each time $t$ the process is in a particular state $s$, and each state generates an observation symbol. However, because the output of the process is usually corrupted by noise, the states are treated as hidden. As an example, HMMs are frequently used in the domain of speech recognition. The underlying hidden states are the actual word symbols that the speaker is uttering, but the only accessible information in the output observation, which is a noisy, continuous speech signal.

Furthermore, probabilities are specified for state transitions (*e.g.*, state A has a 1/3 probability of transitioning to state B, and a 2/3 probability of transitioning to itself) and observation emissions (*e.g.*, state A has a 1/2 probability of emitting observation X and a 1/2 probability of emitting observation Y). A critical assumption of processes modeled with HMMs is that each state only depends on the state before it, and not on any earlier states. This assumption—called the Markov property—is necessary for the operation of the Viterbi algorithm.

Formally, a HMM is defined as a 5-tuple, $(S, O, T, Q, I)$: $S$ is a finite set of (hidden) states that the process could be in; $O$ is the set of observables that the process can emit (of which there could be infinitely many—the observables may come from a continuous space); $T$: $S \times S \rightarrow [0,1]$ is a map, which gives the probability $T(s_1, s_2)$ that the process in state $s_1$ will transition to state $s_2$; $Q$: $S \times O \rightarrow [0,1]$ is a map that gives the probability $Q(s, o)$ that the process in state $s$ will emit observation $o$; and $I$: $S \rightarrow [0,1]$ is a map $I(s)$ that gives the probability that the process will start in a state $s$.

In our setting, we treat the DNA strand as a process that is changing from state to state as it is translocating through the nanopore. The set of states $S$ correspond to the current section of the DNA molecule that the nanopore sequencer is measuring, which might be A, T, C, or G (if the nanopore sequencer is measuring a DNA base), or the non-base E (if the nanopore sequencer is measuring the section of the molecule that is in between two nucleotides). The set of observations consists of the space of DOS histograms—essentially a $B$-dimensional vector space over the real numbers, where $B$ is the number of bins in our histograms. Furthermore, the locality assumptions above justify the tenability of the Markov property in our context.

The probability maps $T$, $Q$, and $I$ can be set in a number of ways. The first way is to define them manually according to one's understanding of the underlying process. The second way is to train the HMM on pre-labeled training data (presumably from experimentation) via an expectation-maximization (EM) algorithm (such as the Baum-Welch algorithm) [25, 26]. EM algorithms are used to estimate the parameters $T$, $Q$, and $I$. In our setting, however, we can tailor the HMM to reflect our knowledge of the observation generation process.

### 2.4.2 Viterbi Algorithm

Given an observation sequence and the particular HMM defined above, we utilize the Viterbi algorithm to discover the most likely sequence of bases to have produced the observation sequence. Below, we briefly summarize the Viterbi algorithm [23].

The Viterbi algorithm uses dynamic programming to efficiently compute the most likely sequence of states to have generated the sequence of observations up to time $t$. Let the given observation sequence be $(\mathbf{D}(t) \mid 1 \leq t \leq T)$. We want to find the sequence of states $(q_1, q_2, \ldots, q_T)$ (where $q_i$ can be any of $\{A,T,C,G,E\}$) such that the conditional probability

$$\Pr[q_1 q_2 \cdots q_T \mid \mathbf{D}(1)\mathbf{D}(2)\cdots \mathbf{D}(T), \lambda] \tag{4}$$

is maximized, where $\lambda$ represents the parameters of the HMM described above. To do so, we first define the quantity

$$\delta_t(i) = \max_{q_1, q_2, \cdots, q_{t-1}} \Pr[q_1, q_2, \cdots, q_t = i \mid \mathbf{D}(1)\mathbf{D}(2)\cdots \mathbf{D}(T), \lambda], \tag{5}$$

*i.e.*, $\delta_t(i)$ represents the highest probability of any sequence of states which ends in state $i \in \{A,T,C,G,E\}$ to have given the sequence of observations up to time $t$. Then by induction, we find that

$$\delta_{t+1}(i) = (\max_i \delta_t(i) T(i,j)) Q(j, \mathbf{D}(t+1)), \tag{6}$$

where $T$ and $E$ are defined as above. We also use an auxiliary variable $\psi_t(j)$ that keeps track of the state that maximized $\delta_t(i)$. We can now specify the complete dynamic program:

1. Initialization

$$\delta_1(i) = I(i)Q(i, \mathbf{D}(t)), \qquad i \in \{A,T,C,G,E\}$$

$$\psi_1(i) = null$$

2. Recursion

$$\delta_t(j) = (\max_i \delta_{t-1}(i)T(i,j))Q(j, \mathbf{D}(t)) \qquad 2 \le t \le T$$

$$j \in \{A,T,C,G,E\}$$

$$\psi_t(j) = \arg\max_{i \in \{A,T,C,G,E\}} [\delta_{t-1}(i)T(i,j)]$$

3. Termination

$$P^* = \max_i \delta_T(i)$$

$$q_T^* = \arg\max_{i \in \{A,T,C,G,E\}} \delta_T(i)$$

4. State sequence backtracking:

$$q_t^* = \psi_{t+1}(q_{t+1}^*), \qquad t = T-1, T-2, \cdots, 1.$$

Then, the final state sequence is given by $(q_1^*, \ldots, q_T^*)$ with probability $P^*$. From this, we discard the $q_t^*$'s that equal $E$ to obtain the most likely nucleotide sequence $(s_1^*, \ldots, s_n^*)$.

# 3  Results and Discussion

To quantitatively evaluate the effectiveness of the proposed sequencing approach, we use simulation data of DNA, for which the ground truth is known.

## 3.1  Simulation Method

We use molecular dynamics (MD) simulations [27] to study the dynamics of DNA molecules in water with counter ions. For selected atomic configurations from the MD simulation trajectory, the electronic density of states are calculated quantum mechanically based on the density functional theory (DFT) [27-29].

Our dataset is generated from simulations of poly(X) molecules composed of two $X$ bases, where $X$ is one of $\{A,T,C,G\}$ (see Fig. 2). All MD simulations are done with the AMBER software [30]. Each simulation begins with energy minimization followed by gradual heating to a temperature of 300 K at atmospheric pressure. Subsequently, each simulation runs for $2 \times 10^6$ steps with a time discretization unit of 1 femtosecond, from which $2 \times 10^3$ atomic configurations (or frames) of 1 picosecond apart are extracted as our working dataset. For each of the 4 systems, 10 equally spaced frames are extracted from the last 100 picoseconds, for a total of 40 frames. For each system, we treat these 10 frames as 10 different configurations of the same molecule, in order to imitate the variability in the poly(X) molecular structure and the noise of an actual DNA nanopore sequencer.

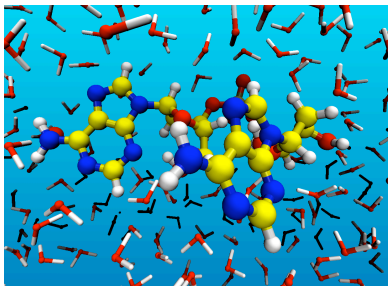

**Fig. 2**. Dimer ssDNA strand consisting of two adenosine molecules, surrounded by water molecules.

Next, we perform DFT calculations [31] to obtain the density of states for each of the 40 configurations. The electronic states are calculated using the projector augmented wave (PAW) method [32], and the generalized gradient approximation (GGA) [33] is used for the exchange-correlation energy. The plane-wave cut-off energies are 30 and 250 Ry for the electronic pseudowave functions and the pseudocharge density, respectively. The energy functional is minimized using an iterative scheme [34]. The Gamma point is used for the Brillouin-zone sampling. Projector functions are generated for the 2s and 2p states of C, N and O, the 3s and 3p states of P, the 1s states of H, and the 2p, 3s, and 3d states of Na. The other electrons in the lower-energy electronic states of each atom are treated with the frozen-core approximation [35].

Figure 3 shows the total DOS's of the systems involving DNA and water. The total DOS largely reflects the electronic structure of surrounding water molecules, and accordingly those corresponding to different DNA bases are not easily distinguishable from each other. The total DOS consists of three peaks at about −3, −9 and −22 eV, which mostly come from the lone-pair 2p state of O, the O-H bonding state, and the s state of O, respectively. The shoulder at ∼ −6 eV originates from some sp hybridization around O. In calculating the electronic structure of DNA, the contribution of water molecules to the DOS is then removed by projecting the DOS on to the subspace spanned by the pseudoatomic orbitals belonging to the DNA [13]. As shown in Fig. 4, we successfully extract the DOS associated with each DNA molecule, which represents recognizable features of the electronic state of DNA.

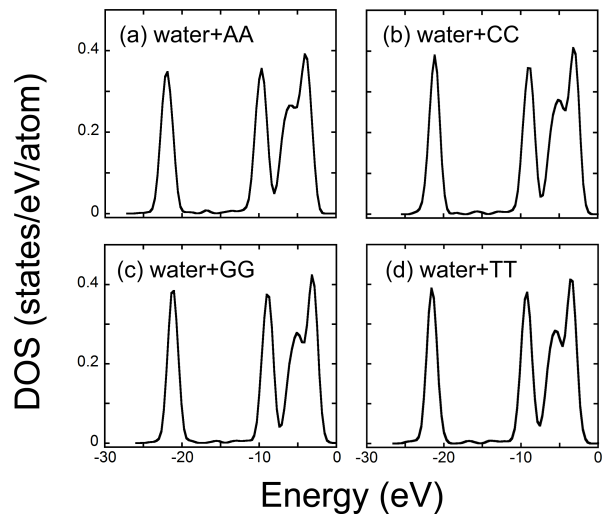

**Fig. 3.** Total electronic density of states of the systems with dimer ssDNA strand and water.

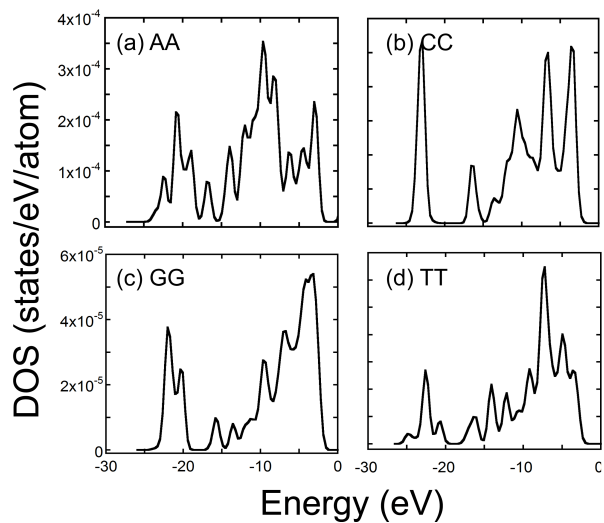

**Fig. 4.** Electronic density of states of dimer ssDNA strand.

To generate DOS histograms, the Kohn-Sham energy eigenvalues of the atomic configurations are partitioned into $B = 30$ bins that equally divide the range $[E_{min}, E_{max}] = [-30 \text{ eV}, 0 \text{ eV}]$. We treat the DOS histograms as $B$ dimensional vectors, where the $i^{th}$ element of the vector corresponds to the $i^{th}$ bin of the histogram. We generated a total of 40 histograms, 10 per each DNA base, out of which we use $4m = 24$ histograms as our training set, with $m = 6$ per each DNA base.

To emulate the DOS time series of a translocating ssDNA strand, we use a simple interpolation scheme: Suppose that our true sequence of bases is $(s_1, s_2, \ldots, s_n)$, where each of the $s_i \in \{\text{A,T,C,G}\}$. For each base $s_i$ in the sequence, the sequencer produces between 1 and 3 observations (DOS histograms), but only one of them corresponds to a direct measurement of the base $s_i$. We call these direct measurements "true" DOS histograms, and the others correspond to some mixed measurement of bases. The DOS histogram for the mixed measurements is computed as linear interpolations between the two true histograms corresponding to $s_i$ and $s_{i+1}$, respectively. Let $\mathbf{d}_i = (a_1, \ldots, a_B)$ and $\mathbf{d}_{i+1} = (b_1, \ldots, b_B)$ be these histograms, where $a_j$ and $b_j$ are the values of the $j^{th}$ bin of the $i^{th}$ and $(i+1)^{th}$ histogram, respectively. Define $\tau(j, p) = (1-p)a_j + pb_j$. If the model chooses to generate 3 observations for $s_i$, for instance, then the first interpolated histogram (after the true histogram) would be $(\tau(1,1/3), \ldots, \tau(B,1/3))$, and the second interpolated histogram would be $(\tau(1,2/3), \ldots, \tau(B,2/3))$. Similarly, if the model choose to generate 2 observations for $s_i$, then the interpolated histogram would be $(\tau(1,1/2), \ldots, \tau(B,1/2))$, which is intuitively the average of the histograms $\mathbf{d}_i$ and $\mathbf{d}_{i+1}$. To simulate the presence of noise, a small amount of Gaussian noise is added to each bin. Here, the noise added to $\mathbf{d}_i$ is drawn from the normal distribution, $N(0,\sigma)$, with zero mean and standard deviation $\sigma = (\text{Var}(a_i)/L)^{1/2}$, where $\mathbf{d}_i = (a_1, \ldots, a_B)$, $\text{Var}(a_i)$ is the variance of $a_i$, and $L$ is some large parameter such as 50. We call the resulting series of observations for the base sequence $\mathbf{D}(t)$, for $1 \leq t \leq T$, where $T \geq n$.

### 3.2 Identifying Single Histograms

We process our sample dataset (described as $\mathbf{H}$ in section 2) through the PCA algorithm. Figure 5 shows the variance of the projection of $\mathbf{H}$ onto the principal components $\mathbf{p}_i$, $i = 1, 2, \ldots, B$. We observe that the first two principal components (indicated by the arrows in Fig. 5) capture the most amount of variance; from the perspective of the rest of the principal components ($i > 2$), $\mathbf{H}$ displays little variation. Intuitively, this indicates that the rest of the PCs describe the noise present in the system. The first two principal components thus "explain" the vast majority of the variance in the data, and accordingly the dimension of the dataset can be reduced to 2 while preserving its underlying structure.

Figure 6 shows the projection space of $\mathbf{H}$, $i.e.$, the projection of DOS vector $\mathbf{d}$ onto the first two principal components. Projection on the first principal component appears to be sufficient to discriminate T and G into distinct clusters, while the second PC is needed to distinguish between A and C. We find that the maximization of variance of $\mathbf{H}$ along the first two PCs corresponds closely to the separation of the histograms according to their nucleotide type. Note that the PCA algorithm is essentially blind to the identities of the histograms, $i.e.$, it "discovered" the separation algorithmically.

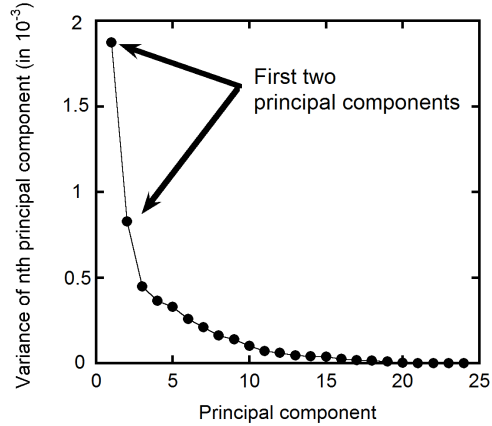

**Fig. 5**. The variance of the dataset **H** with respect to the principal components (PCs), *i.e.* Var($\mathbf{H}\mathbf{p}_i$) is plotted against $i$, for $i = 1, 2, ..., B$. The variances of the first two PCs are indicated by the arrows; much of the variation in the data occurs along the axes $\mathbf{p}_1$ and $\mathbf{p}_2$, while the rest of the PCs describe very little of it.

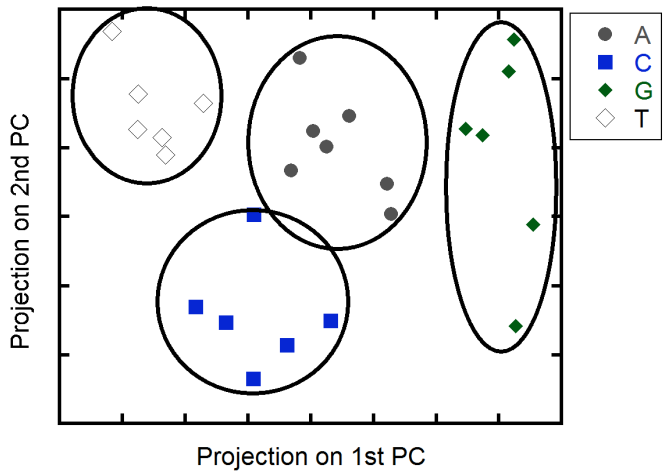

**Fig. 6**. The projection of **H** onto the first two principal components, called the projection space. The nucleotide dataset has been partitioned into disjoint clusters by type (indicated by the circles), indicating that the two principal components have captured the salient features of the density of states histograms.

With any statistical modeling or learning procedure, there is the danger of overfitting, where the features learned do not necessarily describe the underlying patterns and structure of the data, but rather erroneously describe the noise [36]. Thus, the statistic model may learn the training data very well, but have poor predictive performance with new data. We test the results of PCA for its predictive performance in two phases: 1) we present a set of unlabeled histogram samples, and observe the membership probabilities computed for each sample; and then 2) analyze the stability of the clusters in the presence of noise.

We test the results of PCA on the training set by computing $p_A$, $p_T$, $p_C$ and $p_G$ on 12 unlabelled samples (3 for each A, T, C and G). Table 1 shows the results. We identify each histogram $\mathbf{d}$ with $X$ such that $p_X(\mathbf{d}) > 0.5$. Here, 11 out of 12 histograms are correctly identified (91% accuracy), indicating that the features learned by PCA can also classify samples beyond the initial training set.

**Table 1**. The membership probabilities for each of the unknown, sample nucleotide data, against those of randomly generated input vectors. The probabilities greater than 0.5 are bolded.

| Base | # Samples Tested | # Correctly Identified | Avg. $p_A$ | Avg. $p_T$ | Avg. $p_C$ | Avg. $p_G$ |
|------|------------------|------------------------|------------|------------|------------|------------|
| A | 3 | 2 | **0.720** | 0.029 | 0.052 | 0.199 |
| T | 3 | 3 | 0.039 | **0.897** | 0.051 | 0.012 |
| C | 3 | 3 | .038 | 0.034 | **.911** | 0.017 |
| G | 3 | 3 | 0.055 | .011 | 0.014 | **0.920** |

Next, we test the robustness of the clustering in the presence of noise. For this purpose, we take 36 DOS histograms (9 for each nucleotide type) and to each histogram bin add noise sampled from a normal distribution $N(0,\sigma)$ with varying standard deviation $\sigma$. Let $K$ be the average over the maximum amplitude in all DOS vectors in our dataset. A histogram with noise drawn from $N(0,\sigma)$ has a probability of being correctly identified (*i.e.*, if the original histogram is of type $X$, then $p_X(\mathbf{d}) > 0.5$). Figure 7 plots this probability against the parameter $\sigma/K$. The figure shows that an input histogram with noise level $\sigma/K$ up to ~15% will still be more likely than not to be classified correctly. This stability analysis suggests that the features detected by PCA capture the underlying differences between the nucleotides well.

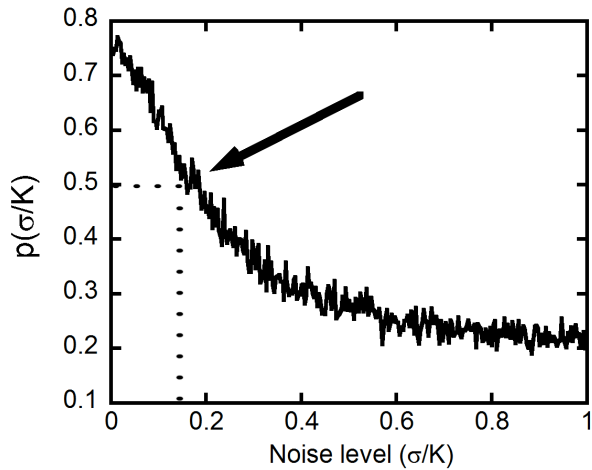

**Fig. 7.** Measurement of the robustness of the clusters with respect to increasing amounts of noise in the input histograms. The vertical axis is the probability that the DOS vector $\mathbf{d}$ of a nucleotide with additive Gaussian noise drawn from $N(0,\sigma)$ is correctly identified (*i.e.*, if the original histogram is of type $X$, then $p_X(\mathbf{d}) > 0.5$), and the horizontal axis is the noise-level parameter $\sigma/K$ , where $K$ is the maximum amplitude over all DOS

vectors in our dataset. The arrow in the graph indicates that a histogram with noise level $\sigma/K$ up to 15% is more likely than not to be correctly identified.

### 3.3 Sequencing a Histogram Time Series

In section 2, we described two approaches to sequencing a given histogram time series $\mathbf{D}(t)$. The first method, Approach 1, is to naively project each $\mathbf{D}(t)$ on the first two PCs found by PCA and then identify $\mathbf{D}(t)$ with nucleotide X such that $p_X(\mathbf{D}(t)) > \tau$, where $\tau$ is some threshold (*e.g.* 0.7). In the event none of the $p_X$ are greater than $\tau$, then the histogram $\mathbf{D}(t)$ is not identified with any nucleotide, and discarded.

Using the interpolation method described in section 3.1, we simulate 5 DOS time series corresponding to 5 randomly generated base sequences. We sequence each time series using the first method. We compare the predicted sequence against the true sequence by measuring the Levenshtein distance between the two strings. The Levenshtein distance between two strings (also known as the "edit distance") measures the minimum number of single character insertions, deletions, substitutions, and transpositions needed to transform one string to another. Table 2 shows the results.

**Table 2**. Sequencing 5 density-of-states histogram time series generated from a hidden nucleotide sequence, using the naive method. **A** is the true nucleotide sequence, whereas **B** is the nucleotide sequence estimated from the histograms. The error for each sequence is calculated as (`Levenshtein distance`) divided by (`True Sequence Length`).

| Sequence Number | True sequence length (A) | Predicted sequence length (B) | Levenshtein distance between A and B | Error |
|---|---|---|---|---|
| 1 | 123 | 208 | 93 | 76% |
| 2 | 195 | 348 | 161 | 83% |
| 3 | 178 | 311 | 143 | 80% |
| 4 | 146 | 249 | 118 | 81% |
| 5 | 120 | 208 | 99 | 83% |

With the same set of 5 generated sequences, we next use the Viterbi algorithm (Approach 2) to predict the most likely sequence of bases to generate the histograms. First, we specify the remaining parameters of the hidden Markov model described in section 2.4.1. We set $T$, $Q$, and $I$ (the transition, observation emission, and initial probabilities, respectively) as follows: For $X_1, X_2 \in \{$A,T,C,G$\}$, $T(X_1, X_2) = 1/3$, $T(E,E) = 1/2$, $T(X_1, E) = 2/3$, and $T(E, X_1) = 1/2$. One can verify that these state transition probabilities are consistent with the model of observation generation described in section 3.1. For the observation emission probabilities, we take advantage of the probability functions $p_x$ that we defined above, with a slight modification: We have to introduce the probability that the HMM, in state $E$, will emit a given DOS observation vector $\mathbf{o}$. We will define it as follows:

$$p_E(\mathbf{o}) = (1 - p_A(\mathbf{o}))(1 - p_T(\mathbf{o}))(1 - p_C(\mathbf{o}))(1 - p_G(\mathbf{o})) \, . \tag{2}$$

However, the emission probabilities must be normalized, so for $X \in \{A,T,C,G,E\}$:

$$Q(X,\mathbf{o}) = \frac{p_X(\mathbf{o})}{\sum_{N \in \{A,T,C,G,E\}} p_N(\mathbf{o})} \, . \tag{3}$$

Hence for a fixed $\mathbf{o}$, $\sum_X Q(X,\mathbf{o}) = 1$. Finally, the initial probability is $I(X) = 1/5$.

Table 3 shows the results from Approach 2. The error rate has been reduced four-fold (average 21%) compared with that of Approach 1.

**Table 3**. Sequencing the same 5 time-series as in Table 2, but using the Viterbi algorithm.

| Sequence Number | True sequence length (A) | Predicted sequence length (B) | Levenshtein distance between A and B | Error |
|:---:|:---:|:---:|:---:|:---:|
| 1 | 123 | 117 | 28 | 23% |
| 2 | 195 | 189 | 42 | 22% |
| 3 | 178 | 176 | 33 | 19% |
| 4 | 146 | 148 | 30 | 21% |
| 5 | 120 | 119 | 28 | 23% |

The above results indicate that modeling the DNA sequences as Markov chains allowed for much more accurate sequencing via the Viterbi algorithm, as compared to the naive approach. With Approach 1, the predicted sequences have a huge excess of nucleotide symbols that are a byproduct of the interpolated histograms. The Viterbi algorithm is able to recognize that many of the histograms corresponded to interstitial regions.

Though 21% error is quite far from acceptable error rates (reliable human genome sequencing requires an error rate of at most of 1/100,000 base pairs [37]), our results at least demonstrate the benefit of using HMM in DNA sequencing. The results of using the naive approach indicate that simply using PCA alone to identify DNA bases is unworkable, and added knowledge about the nanopore sequencing process must be utilized. Here, with an extremely simple HMM, the Viterbi algorithm was able to significantly improve the accuracy of sequencing. With a more sophisticated HMM, the Viterbi algorithm should be able to sequence DOS time series with much greater accuracy. For example, Boufounos *et al.* have suggested such an alternative topology for the HMM, involving not only the 4 base types but 16 additional states representing the transitions between nucleotide types (*e.g.* AA, AT, AC, *etc.*) [38]. With training data for these nucleotide transition areas, the Viterbi algorithm should be able to recognize the non-nucleotide DOS more easily. Another possibility would be to use trimer training data. This will allow for a more nuanced HMM for DNA sequencing, where the states are not simply one of $\{A,T,C,G,E\}$, but would encompass all possible codons: AAA, AAT, AAC, *etc*. Such expanded HMM would capture more of the subtleties involved in the input signal.

# 4 Conclusions

In summary, we have presented theoretical results for methods of classifying both single DNA nucleotide molecules and a sequence of DNA bases from the electronic density of states of the nucleotides. Our results indicate that the density of states profiles of the DNA nucleotides may have enough information to distinguish between the base identities, and that the combination of principal component analysis and the Viterbi algorithm can extract this information to solve the nucleotide identification problem.

With principal component analysis, we were able to reduce the dimensionality of the histogram space into a two-dimensional "projection space," which partitioned the training set of histograms into distinct clusters. We demonstrated that the partitioning of the projection space predicts the identity of an unlabeled histogram very well, and that the clusters are stable with respect to moderate levels of Gaussian noise (density of states histograms are correctly identified the majority of the time even where the added noise is 15% of the maximum histogram amplitude).

We then compared two approaches to sequencing a time series of histograms (as would be generated by a nanopore sequencer device). The first was a naive approach of identifying each histogram in the time series individually via the clusters found with principal component analysis. The second approach was to model the time series of histograms as the product of a hidden Markov model process and use the Viterbi algorithm to find the most likely sequence of bases that generated the histograms.

Our numerical experiments demonstrated that the Viterbi algorithm performs vastly superior to the naive approach. Simply identifying each histogram $\mathbf{D}(t)$ with base $X$ such that $p_X(\mathbf{D}(t)) > \tau$ for some threshold $\tau$ was producing an average 80% error rate on a test dataset, whereas the Viterbi algorithm, on the same set of data, was able to achieve an average 21% error rate (with the errors measured with the Levenshtein metric).

We believe that our proposed methods are viable in experiments, and the results and ideas presented here may serve as a step closer towards the realization of the Ultra Low Cost Sequencer.

## Acknowledgements

This work was supported by NSF-PetaApps and NSF-EMT grants. Simulations were performed on the 11,664-processor Linux cluster at USC's High Performance Computing Facility and on the 2,048-processor Linux cluster at our Collaboratory for Advanced Computing and Simulations.